\documentclass[floats,prd,aps,showpacs,amssymb,nofootinbib]{revtex4-1}
\usepackage{amssymb}
\usepackage{graphics}
\usepackage{amsmath}
\usepackage{amsfonts}
\usepackage{bm}
\usepackage[]{latexsym}
\RequirePackage[dvipsnames,usenames]{color}

\newcommand{\be}{\begin{equation}}
\newcommand{\ee}{\end{equation}}
\newcommand{\bea}{\begin{eqnarray}}
\newcommand{\eea}{\end{eqnarray}}
\newcommand{\brr}{\begin{array}}
\newcommand{\err}{\end{array}}
\newcommand{\bit}{\begin{itemize}}\newcommand{\eit}{\end{itemize}}
\newcommand{\ben}{\begin{enumerate}}\newcommand{\een}{\end{enumerate}}

\newcommand{\ba}{\begin{array}}
\newcommand{\ea}{\end{array}}

\def\lf{\left}

\def\ri{\right}
\def\al{\alpha}

\def\1{{_{1}}}\def\2{{_{2}}}
\def\bx{{\bf {x}}}

\def\noHe0{:\;\!\!\;\!\!:H_e(0):\;\!\!\;\!\!:}
\def\noHm0{:\;\!\!\;\!\!:H_\mu(0):\;\!\!\;\!\!:}

\def\lf{\left}

\def\ri{\right}

\def\al{\alpha}

\def\1{{_{1}}}\def\2{{_{2}}}

\def\uuu{U}

\newcommand{\lp}{\ell_{\rm p}}
\newcommand{\mpl}{m_{\rm p}}
\newcommand{\gn}{G_{\rm N}}

\newcommand{\Ep}{\mathcal{E}_{\rm p}}
\begin{document}
\title{Modified Unruh effect from Generalized Uncertainty Principle}%
\author{Fabio Scardigli\footnote{corresponding author}}
\email{fabio@phys.ntu.edu.tw}
\affiliation{Dipartimento di Matematica, Politecnico di Milano, Piazza Leonardo da Vinci 32, 20133~Milano, Italy}
\affiliation{Institute-Lorentz for Theoretical Physics, Leiden University, P.O.~Box 9506, Leiden, The Netherlands}

\author{Massimo Blasone}
\email{blasone@sa.infn.it}
\affiliation{Dipartimento di Fisica, Universit\`a di Salerno, 84084~Fisciano (SA), Italy}
\affiliation{INFN, Gruppo Collegato di Salerno}
\author{Roberto Casadio}
\email{casadio@bo.infn.it}
\affiliation{Dipartimento di Fisica e Astronomia, Alma Mater Universit\`a di Bologna, via Irnerio~46,
40126 Bologna, Italy}
\affiliation
{INFN, Sezione di Bologna, viale Bert~Pichat~6/2, 40127~Bologna, Italy}
\author{Gaetano Luciano}
\email{gluciano@sa.infn.it}
\affiliation{Dipartimento di Fisica, Universit\`a di Salerno, 84084~Fisciano (SA), Italy}
\affiliation{INFN, Gruppo Collegato di Salerno}
%
%
%
  \def\be{\begin{equation}}
\def\ee{\end{equation}}
\def\al{\alpha}
\def\bea{\begin{eqnarray}}
\def\eea{\end{eqnarray}}
\begin{abstract}
We consider a generalized uncertainty principle (GUP) corresponding to a deformation of the fundamental commutator obtained by adding a term quadratic in the momentum.
From this GUP, we compute corrections to the Unruh effect and related Unruh temperature, by first following a heuristic derivation, and then a more standard field theoretic calculation.
In the limit of small deformations, we recover the thermal character of the Unruh radiation.
Corrections to the temperature at first order in the deforming parameter are compared for the two approaches, and found to be in agreement as for the dependence on the cubic power of the acceleration of the reference frame.
The dependence of the shifted temperature on the frequency is also pointed out and discussed.
\end{abstract}
 \vskip -1.0 truecm
\maketitle
\section{Introduction}
In the last thirty years, many studies have converged on the idea that the Heisenberg uncertainty
principle (HUP)~\cite{Heisenberg} should be modified when gravitation is taken into account.
In microphysics, gravity is usually neglected on the ground of its weakness,
when compared with the other fundamental interactions.
However, this argument should not apply when one wants to address fundamental questions in Nature. In this perspective, gravity should be included, especially when we discuss the formulation of a fundamental principle like the Heisenberg's one.
And in fact, gravitation has always played a pivotal role in the generalization of the HUP,
from the early attempts~\cite{GUPearly}, to the more recent proposals, like those in string theory, loop quantum gravity, deformed special relativity, non-commutative geometry, and studies of black hole physics~\cite{VenezGrossMende,MM,FS,Adler2,CGS,SC2013}.
\par
A possible way for this generalization is to reconsider the well-known classical argument of the
Heisenberg microscope~\cite{Heisenberg}. The size $\delta x$ of the smallest detail of an object, theoretically detectable with a beam of photons of energy $E$, is roughly given by (assuming the dispersion relation $E=p$)~\footnote{We shall always
work with $c=1$, but explicitly show the Newton constant $\gn$ and the Planck constant 
$\hbar$. The Planck length is defined as $\lp=\sqrt{\gn\,\hbar/c^3}\simeq 10^{-35}\,$m, the Planck energy as $\Ep\,\lp = \hbar\, c/2$, and the Planck mass
as $\mpl=\Ep/c^2\simeq 10^{-8}\,$kg, so that $\lp=2\gn\,\mpl$ and $2\,\lp\,\mpl= \hbar$. The Boltzmann constant $k_{\rm B}$ will be shown explicitly, unless otherwise stated.}
\be
\delta x
\simeq
\frac{\hbar}{2\, E}
\ ,
\label{HS}
\ee
since increasingly large energies are required to explore decreasingly small details.
In its original formulation, Heisenberg's gedanken experiment ignores gravity.
However, gedanken experiments involving formation of gravitational instabilities
in high energy scatterings of strings~\cite{VenezGrossMende}, or gedanken experiments taking
into account the possible formation, in high energy scatterings, of micro black holes with a
gravitational radius $R_S=R_S(E)$ proportional to the (centre-of-mass) scattering energy $E$ 
(see Ref.\cite{FS}), suggest that the usual uncertainty relation should be modified as
\be
\delta x
\simeq
\frac{\hbar}{2\, E}
+
\beta\, R_S(E)
\ ,
\ee
where $\beta$ is a dimensionless parameter.
Recalling that $R_S\simeq 2\,\gn\, E = 2\, \lp^2\, E/\hbar$, we can write
\be
\delta x
\simeq
\frac{\hbar}{2\, E}
+
2\beta\, \lp^2\frac{E}{\hbar}
=
\lp
\left(
\frac{\mpl}{E}
+
\beta\, \frac{E}{\mpl}
\right)
\ .
\label{He}
\ee
This kind of modification of the uncertainty principle was also proposed in Ref.~\cite{Adler2}.
\par
The dimensionless deforming parameter $\beta$ is not (in principle) fixed by the theory, although it is generally assumed to be of order one.
This happens, in particular, in some models of string theory (see again for instance Ref.~\cite{VenezGrossMende}), and has been confirmed by an explicit calculation in Ref.~\cite{SLV}.
However, many studies have appeared in literature, with the aim to set bounds on $\beta$
(see, for instance, Refs.~\cite{brau}).
\par
The relation~(\ref{He}) can be recast in the form of an uncertainty relation, namely a deformation of the standard HUP, usually referred to as Generalized Uncertainty Principle (GUP),
\be
\Delta x\, \Delta p
\geq
\frac{\hbar}{2}
\left[1
+\beta
\left(\frac{\Delta p}{\mpl}\right)^2
\right]
\ .
\label{gup}
\ee
For mirror-symmetric states (with $\langle \hat{p} \rangle = 0$), the inequality (\ref{gup}) is equivalent to the commutator
\be
\left[\hat{x},\hat{p}\right]
=
i \hbar \left[
1
+\beta
\left(\frac{\hat{p}}{\mpl} \right)^2 \right]
\ ,
\label{gupcomm}
\ee
since $\Delta x\, \Delta p \geq (1/2)\left|\langle [\hat{x},\hat{p}] \rangle\right|$.
Vice-versa, commutator (\ref{gupcomm}) implies inequality (\ref{gup}) for any state.
The GUP is widely studied in the context of quantum mechanics~\cite{Pedram},
quantum field theory~\cite{Husain:2012im,Majhi:2013koa}, quantum gravity~\cite{Hossain:2010wy},
and for various deformations of the quantization rules~\cite{Hossain:2010wy, Jizba:2009qf}.
The above $\beta$-deformed commutator~(\ref{gupcomm}) will be the starting point of the present
investigation. 
In what follows, using~(\ref{gupcomm}), we shall describe the Unruh effect (known also as Fulling-Davies-Unruh effect~\cite{Fulling:1972md,davies,Unruh:1976db}), thereby calculating corrections to the Unruh temperature to first order in $\beta$. A direct derivation of the Unruh effect from the HUP has been given in Ref.~\cite{FS9506}. On the other hand, the necessity of this effect for the internal consistency of QFT has been confirmed by arguments based both on general covariance~\cite{Matsas:1999jx3} and thermodynamic~\cite{Becattini:2017ljh}. Moreover, non-trivial modifications to the Unruh spectrum have been recently pointed out also in different contexts, for instance, it has been shown that flavor mixing does spoil its thermal character~\cite{Blasone:2017nbf, Blasone:2018byx}, thus opening new stimulating scenarios.
\section{Heuristic derivation of Unruh Effect from uncertainty relations}
In this section we derive the Unruh temperature~\cite{Unruh:1976db} starting directly
from the HUP. Simple classical physics relations will be used together with the quantum principle, following closely Ref.~\cite{FS9506} (see also the recent Ref.~\cite{gine}).
This procedure will then allow us to estimate what kind of corrections are induced by a GUP.
\par
Let us consider some elementary particles, for example electrons, kept at rest in an uniformly
accelerated frame.
The kinetic energy acquired by each of these particles while the accelerated frame moves a
distance $\delta x$ will be given by
\be
E_k
=
m\,a\,\delta x
\ ,
\ee
where $m$ is the mass of the particle and $a$ the acceleration of the frame, and therefore
of the particle.
Now, suppose this energy is sufficient to create $N$ pairs of the same kind of particles from the quantum
vacuum.
Namely, we set
\be
E_k
\simeq
2\,N\, m
\ ,
\ee
and find that the distance along which each particle must be accelerated in order to create $N$ pairs is
\be
\delta x
\simeq
2\,\frac{N}{a}
\ .
\label{dx}
\ee
The original particles and the pairs created in this way are localized inside a spatial region of width $\delta x$,
therefore the fluctuation in energy of each single particle is
\be
\delta E
\simeq
\frac{\hbar}{2\, \delta x}
\simeq
\frac{\hbar\, a}{4\, N}
\ .
\ee
If we interpret this fluctuation as a classical thermal agitation of the particles, we can write
\be
\frac{3}{2}\,k_{\rm B}\,T
\simeq
\delta E
\simeq
\frac{\hbar\, a}{4\,N}
\ ,
\label{dE}
\ee
or
\be
T
=
\frac{\hbar\, a}{6\,N\,k_{\rm B}}
\ .
\ee
On comparing with the well-known Unruh's temperature~\cite{Unruh:1976db},
\be
T_{\rm U}
=
\frac{\hbar\, a}{2\,\pi\, k_{\rm B}}
\ ,
\label{Tu}
\ee
we can set the arbitrary parameter $N$ and obtain an effective number of pairs $N=\pi/3\simeq 1$.
\par
Now we repeat the same argument using the GUP.
Upon replacing Eq.~\eqref{dx} into Eq.~\eqref{He}, and interpreting the energy fluctuation $\delta E$
in terms of a classical thermal bath, we find
\be
2\,\frac{N}{a}
\simeq
\frac{\hbar}{3\, k_{\rm B}\, T}
+
\beta\, \lp^2\, \frac{3\, k_{\rm B}\, T}{\hbar}
\ .
\label{approx}
\ee
Requiring that the $T$ equals the Unruh temperature~\eqref{Tu} for $\beta \to 0$ again fixes $N=\pi/3\simeq 1$,
and we finally obtain
\be
\frac{2\,\pi}{a}
\simeq
\frac{\hbar}{k_{\rm B}\, T}
+
9\beta\, \lp^2\, \frac{k_{\rm B}\, T}{\hbar}
=
\lp\left(
\frac{2\mpl}{k_{\rm B}\,T}
+
9\,\beta\,\frac{k_{\rm B}\, T}{2\mpl}
\right)
\ .
\label{acctemp}
\ee
This relation can be easily inverted for $T=T(a)$.
However, it is reasonable to assume that $\beta\,k_{\rm B} T/\mpl\sim \beta \,m/\mpl$ is very small for any fundamental particle with
$m\ll \mpl$.
We can therefore expand in $\beta\,m/\mpl$ and find
\be
T
\simeq
T_{\rm U}
\left(1
+
\frac{9\,\beta}{4}\,\frac{\lp^2\,a^2}{\pi^2}
\right)
=
T_{\rm U}
\left[1
+
\frac{9\,\beta}{4}
\left(\frac{k_{\rm B}\,T_{\rm U}}{\mpl}
\right)^2
\right]
\ .
\label{newTHeuristic}
\ee
We also notice an interesting physical property suggested by Eq.~(\ref{acctemp}), that is, by the GUP. In order to maintain the inverted relation $T=T(a)$ physically meaningful (i.e. the temperature must be a real number), there will be a maximal value for the acceleration, namely
\be
a
\lesssim
\frac{\pi}{3 \, \sqrt{\beta}\, \lp}
\ ,
\ee
and a corresponding maximal value for the Unruh-Davies temperature,
\be
k_{\rm B}\,T_{\rm U}
\lesssim
\frac{\mpl}{3\,\sqrt{\beta}}
\ .
\ee
These ideas and estimates naturally make contact with those reported, for example, in Refs.~\cite{caianiello}.
\section{Quantization of a massive scalar field in accelerated frame}
\label{Quantization}
In this Section we briefly review the quantization of a massive scalar field for an accelerated observer. This will serve as a basis for the analysis of Section~\ref{GUP}, where the deformation of the algebra discussed above is implemented.
For the sake of simplicity, we will work in $1+1$-dimensions, using the Minkowski metric with the
conventional signature $ds^2= \eta_{\mu\nu}\,dx^\mu\,dx^\nu=dt^2-dx^2$. In this Section we set $\hbar=c=k_{\rm B}=1$, unless otherwise explicitly stated.
\subsection{Minkowski spacetime}
For an inertial observer, the scalar field in the usual plane-wave representation reads
\begin{equation}
\phi(\bx)
=
\int dk\,
\left[
a_{k}\,\uuu_{k}(\bx)
+
a_{k}^\dagger\,\uuu_{k}^{*}(\bx)
\right]
\ ,
\label{eqn:planewavexpans}
\end{equation}
where $\bx\equiv\{t, x\}$ denotes the set of Minkowski coordinates.
The positive frequency plane-waves of momentum $k$ are given by
\begin{equation}
U_{k}(\bx)
=
{\left(4\,\pi\,\omega_{k}\right)}^{-\frac{1}{2}}\,
e^{i\left(k\,x-\omega_{k}\, t\right)}
\ ,
\label{eqn:modes}
\end{equation}
where $\omega_{k}=\sqrt{m^2+k^2}$, $m$ being the mass of the field.
Within the framework of canonical quantum field theory (QFT), the annihilation and creation 
operators for Minkowski quanta, to wit $a_{k}$ and $a^\dagger_{k}$, satisfy the standard
commutation relation
\begin{equation}
\left[a_k, a_{k'}^\dagger\right]
=
\delta(k-k')
\ ,
\label{eqn:commutcanon}
\end{equation}
with all other commutators vanishing.
The ordinary Minkowski vacuum is accordingly defined by $a_k\,|0_M\rangle=0$ for all modes $k$.
\par
As a tool for extending this quantization scheme to an accelerated observer, let us now introduce the less familiar
Lorentz-boost eigenfunctions~\cite{Blasone:2017nbf}.
Boost modes are related to the plane-waves in Eq.~(\ref{eqn:modes}) by
\begin{equation}
\widetilde{U}_{\Omega}^{(\sigma)}(\bx)
=
\int dk\, p_\Omega^{(\sigma)*}(k)\,U_{k}(\bx)
\ ,
\label{eqn:Uwidetildemodes}
\end{equation}
where
\begin{equation}
\label{eqn:p}
p_\Omega^{(\sigma)}(k)
=
\frac{1}{\sqrt{2\,\pi\,\omega_{k}}}\,
\left(\frac{\omega_{k}+k}{\omega_{k}-k}\right)^{i\,\sigma\,\Omega/2},
\qquad
\sigma
=
\pm
\ ,
\quad
0<\Omega<\infty
\ .
\end{equation}
The physical meaning of the quantum numbers $\Omega$ and $\sigma$ will be discussed in the next Section.
In terms of the modes~(\ref{eqn:Uwidetildemodes}), the spectral representation of the field operator can be written as \footnote{Note that, although the plane-wave field expansion in Eq.~(\ref{eqn:planewavexpans}) applies to the whole of the Minkowski space-time, the representation Eq.~(\ref{eqn:expansionfieldboost}) in terms of boost-modes does hold only in the Rindler manifold $x>|t|\hspace{0.4mm}\cup\hspace{0.4mm} x<-|t|$ (see Fig.1). A globally well-defined expansion can be obtained by analytically continuing the modes Eq.~(\ref{eqn:Uwidetildemodes}) across the null asymptotes $x=\pm\, t$ (see Ref.~\cite{Gerlach}). For our purposes, however, it is enough to consider the definition Eq.~(\ref{eqn:Uwidetildemodes}) of boost modes.}
\begin{equation}
\phi(\bx)
=
\int_{0}^{+\infty}
d\Omega\, 
\sum_{\sigma}
\left[d_{\Omega}^{(\sigma)}\,\widetilde{U}_{\Omega}^{(\sigma)}(\bx)
+
d_{\Omega}^{(\sigma)\dagger}\,\widetilde{U}_{\Omega}^{(\sigma)*}(\bx)
\right]
\ .
\label{eqn:expansionfieldboost}
\end{equation}
\par
It is easy to prove that the two quantum constructions introduced above are equivalent to each other.
For this purpose, let us equate the field-expansions~(\ref{eqn:planewavexpans})
and~(\ref{eqn:expansionfieldboost}) on a space-like hypersurface.
By using Eq.~(\ref{eqn:Uwidetildemodes}), it follows that
\begin{equation}
\label{eqn:operat-d}
d_{\Omega}^{(\sigma)}
=
\int dk\,
p_\Omega^{(\sigma)}(k)\,a_{k}
\ .
\end{equation}
Since the operators $d_{\Omega}^{\,(\sigma)}$ are linear combinations of the Minkowski
annihilators $a_{k}$ alone, they also annihilate the Minkowski vacuum $|0_M\rangle$.
Moreover, by exploiting the completeness and orthonormality of the set of functions 
$\lf\{p_\Omega^{(\sigma)}\ri\}$ (see Ref.~\cite{Takagi:1986kn}), it can be shown
that the transformation~(\ref{eqn:operat-d}) is canonical, so that
\begin{equation}
\label{eqn:d-commut}
\left[d_{\Omega}^{(\sigma)},d_{\Omega'}^{\,(\sigma')\dagger}\right]
=
\delta_{\sigma\sigma'}\,\delta(\Omega-\Omega')
\ ,
\end{equation}
with all other commutators vanishing.
Eqs.~(\ref{eqn:operat-d}) and (\ref{eqn:d-commut}) allow us to interpret also the 
$d_{\Omega}^{\,(\sigma)}$ as annihilation operators of Minkowski quanta. This implies that the
field-expansions Eqs.~(\ref{eqn:planewavexpans}) and~(\ref{eqn:expansionfieldboost}) can be used equivalently within the framework of canonical quantization in Minkowski space-time.
For our purposes, in what follows it will be convenient to employ the latter.
\subsection{Rindler space-time}
\begin{figure}[t]
\resizebox{8cm}{!}{\includegraphics{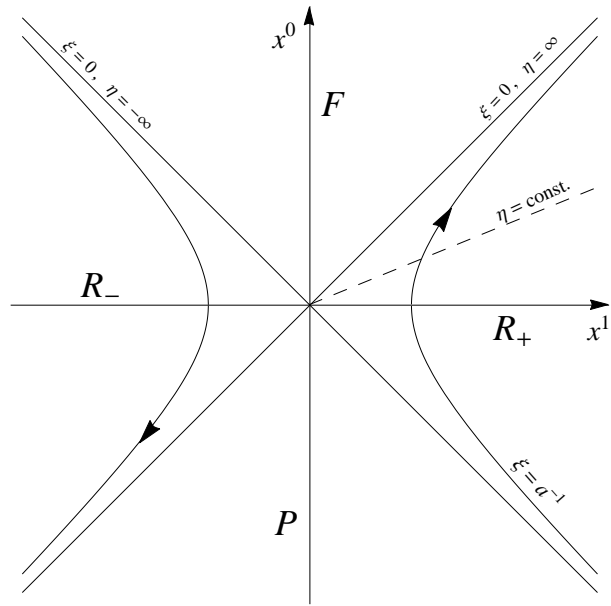}}
\caption{\small{The proper coordinate system of a uniformly accelerated observer in the Minkowski spacetime.
The branch of hyperbola $\xi=a^{-1}$ represents the worldline of an observer with proper acceleration $a$.}}
\label{figure:Rindler}
\end{figure}

The foregoing discussion applies to inertial observers in Minkowski space-time.
In order to investigate GUP effects on the Unruh radiation~\cite{Unruh:1976db},
let us now review the Rindler-Fulling field-quantization in a uniformly accelerating frame~\cite{Fulling:1972md}.
By introducing the usual Rindler coordinates $\{\eta,\xi\}$, in place of $\{t, x\}$, we have 
\begin{equation}
\label{eqn:rindlercoordinates}
t
=
\xi\,\sinh\eta
\ ,
\quad
x=
\xi\,\cosh\eta, \qquad -\infty\,<\,\eta,\xi\,<\,\infty,
\end{equation}
and the Minkowski line element takes the well-known form
\begin{equation}
ds^2
=
dt^2-dx^2
=
\xi^2\,d\eta^2
-d\xi^2
\ .
\label{eqn:lineelement}
\end{equation}
As $\xi$ and $\eta$ range from $-\infty$ to $\infty$, the Rindler coordinates cover only two sections of Minkowski space-time, specifically the right wedge $R_+=\lf\{{\bf x}\,|\,x>|t|\ri\}$ for $\xi>0$,
and the left wedge $R_-=\lf\{{\bf x}\,|\,x<-|t|\ri\}$ for $\xi<0$ (see Fig.~\ref{figure:Rindler}).
Since the components of the  metric in these coordinates do not depend on $\eta$,
Eq.~(\ref{eqn:lineelement}) describes a static spacetime with Killing vector ${\bf B}=\partial_\eta$.
\par
The worldline of a uniformly accelerated (Rindler) observer with proper acceleration $|a|$ is given by
\begin{equation}
\xi(\tau)
=
\mathrm{const}
\equiv
a^{-1}
\ ,
\label{eqn:lineauniverso}
\end{equation}
where $\tau=\eta/a$ is the proper time along the accelerated trajectory (restoring $c\neq 1$, one has $\xi(\tau)=c^2/a$, and $\tau=\eta c/a$; that is, $\eta=a\tau/c$ is dimensionless).
This is a branch of hyperbola in the $(t,x)$ plane, whose null asymptotes $t=\pm x$ act respectively as future and past event horizons for the Rindler observer.
\par
Because of the non-trivial structure of  Rindler space-time, the wedges $R_\pm$ are causally
disconnected from each other~\cite{Rindler:1966zz}.
The positive frequency solutions of the Klein-Gordon equation in Rindler coordinates thus take the
form~\footnote{In what follows, the set of Rindler coordinates $\{\eta, \xi\}$ will be also denoted by $\bx$; therefore, such a symbol will refer to a space-time point, rather than its representative in a particular coordinate system.}
\begin{equation}
u_\Omega^{(\sigma)}(\bx)
=
N_\Omega\,
\theta(\sigma\,\xi)\,
K_{i\,\Omega}^{(\sigma)}(m\,\xi)\,
e^{-i\,\sigma\,\Omega\,\eta}
\ ,
\label{eqn:rindlermodes}
\end{equation}
where $\Omega$ is the Rindler frequency with respect to the time $\eta$~\footnote{Using $c\neq1$ for sake of clarity, the proper frequency $\omega$ measured by a Rindler observer is obtained from $\omega\,\tau = \omega(\eta c/a)=(\omega c/a)\eta\equiv\Omega\,\eta$.}, $\sigma=\pm$ refers to the right/left wedges 
$R_\pm$ and $K_{i\,\Omega}$ is the modified Bessel function of the second kind.
In this context, we do not need to specify the normalization factor $N_\Omega$
(for more details, see Ref.~\cite{Takagi:1986kn}).
Furthermore, the Heaviside function $\theta(\sigma\,\xi)$ was inserted into Eq.~(\ref{eqn:rindlermodes}) in order to constrain the Rindler modes to only one of the two disconnected wedges $R_\pm$.
\par
Using Eq.~(\ref{eqn:rindlermodes}), we can now expand the scalar field operator in the Rindler space-time as follows
\begin{equation}
\phi(\bx)
=
\int_{0}^{+\infty}
d\Omega\,
\sum_\sigma\,
\left[b^{(\sigma)}_{\Omega}\,u_{\Omega}^{(\sigma)}(\bx)
+b^{(\sigma)\dagger}_{\Omega}\,u_{\Omega}^{(\sigma)*}(\bx)
\right]
\ ,
\label{eqn:espanrind}
\end{equation}
where the ladder operators $b^{\,(\sigma)}_\Omega$ and ${b^{\,(\sigma)\dagger}_{\Omega}}$
are assumed to satisfy the canonical commutation relations
\begin{equation}
\left[b_\Omega^{(\sigma)}, b_{\Omega'}^{( \sigma')\dagger}\right]
=
\delta_{\sigma\sigma'}\,
\delta(\Omega-\Omega')
\ ,
\label{eqn:commutcanon2}
\end{equation}
with all other commutators vanishing.
The Rindler vacuum is accordingly defined by $b_\Omega^{(\sigma)}|0_{\rm R}\rangle = 0$, for all 
values of $\sigma$ and $\Omega$.
\par
The connection between the two quantization schemes, for inertial and accelerated observers, can now be investigated in detail. Specifically, we compare the 
field-expansions~(\ref{eqn:expansionfieldboost}) and (\ref{eqn:espanrind}) on a spacelike hypersurface $\Sigma$ lying in the Rindler manifold $R_\pm$.
A straightforward calculation leads to the following Bogoliubov 
transformation~\cite{Takagi:1986kn}
\begin{equation}
b^{(\sigma)}_{\Omega}
=
\left[1+\mathcal{N}(\Omega)\right]^{1/2}\,
d_{\Omega}^{(\sigma)}
+
\mathcal{N}(\Omega)^{1/2}\,d_{\Omega}^{(-\sigma)\dagger}
\ ,
\label{eqn:bogotransform}
\end{equation}
where 
\begin{equation}
\label{eqn:B-Edist}
\mathcal{N}(\Omega)
=
\frac{1}{e^{2\,\pi\,\Omega}-1}
\end{equation}
is the Bose-Einstein distribution.
Using Eq.~(\ref{eqn:bogotransform}), we can now calculate the spectrum of Rindler quanta in the Minkowski vacuum $|0_{\rm M}\rangle$,
\begin{equation}
\langle0_{\rm M}|\,b_{\Omega}^{(\sigma)\dagger}\,b_{\Omega'}^{ (\sigma')}\,|0_{\rm M}\rangle
=
\mathcal{N}(\Omega)\,\delta_{\sigma\sigma'}\,\delta(\Omega-\Omega')
\ .
\label{eqn:aspectval}
\end{equation}
It then follows that a uniformly accelerated observer perceives the Minkowski vacuum as a thermal bath
of Rindler quanta with a temperature proportional to the acceleration (Unruh effect~\cite{Unruh:1976db}).
Restoring our standard units ($c=1$, $\hbar\neq 1$, $k_{\rm B}\neq 1$), we can in fact write 
\be
2\,\pi\,\Omega
=
\frac{2\pi}{a}\,a\,\Omega
=
\frac{\hbar \,a\, \Omega}{k_{\rm B}\,T_{\rm U}}
=
\frac{\hbar\,\omega}{k_{\rm B}\,T_{\rm U}}
\ ,
\ee
where $\omega=a\,\Omega$ is the frequency measured by the Rindler observer and $T_{\rm U}$
the Unruh temperature~\eqref{Tu}.
\section{GUP and modified Unruh temperature}

\label{GUP}
In the previous Section, the Unruh temperature~(\ref{Tu}) has just been re-derived within the framework of the canonical QFT. At this stage, one may wonder how such a result gets modified when starting from the GUP commutator in Eq.~(\ref{gupcomm}).
To answer this question, an intermediate step concerning the effects of GUP on a quantum one-dimensional harmonic oscillator turns out to be useful.
In this context, we note that the ladder operators $A$ and $A^\dagger$ of the deformed algebra for the one-dimensional harmonic oscillator are linked to $\hat{x}=\hat{x}^\dagger$ and $\hat{p}=\hat{p}^\dagger$ by the usual relations
\be
	A = \frac{1}{\sqrt{2m\hbar\omega}}(m\omega\hat{x} + i \hat{p}), \qquad
	A^\dagger = \frac{1}{\sqrt{2m\hbar\omega}}(m\omega\hat{x} - i \hat{p})\,,
\ee
and their inverses
\be
	\hat{x} = \sqrt{\frac{\hbar}{2m\omega}}(A^\dagger + A), \qquad
	\hat{p} = i\sqrt{\frac{m\hbar\omega}{2}}(A^\dagger - A) \,.
\ee
It is then easy to see that 
\be
[A,A^\dagger]\,=\,\frac{1}{i\hbar}\,[\hat x, \hat p]
\ee
and, due to the modified commutator (\ref{gupcomm}) between 
$\hat x$ and $\hat p$, the deformed algebra for the one-dimensional harmonic oscillator should be written as
\begin{equation}
\label{eqn:oscillharm}
\left[A, A^\dagger\right]
=
\frac{1}{1-\alpha}
\left[1
-\alpha
\left(A^\dagger\, A^\dagger
+A\,A
-2\,A^\dagger\, A
\right)
\right]
\ ,
\end{equation}
where 
\be
\alpha=\beta\,\frac{m\,\hbar \omega}{2\,\mpl^2}
\ ,
\ee
with $m$ and $\omega$ being the mass and frequency of the harmonic oscillator, respectively.
The modified quantization rules~(\ref{eqn:oscillharm}) can be now extended in a natural way to a scalar field in the plane-wave representation, if we consider that, for a given momentum $k$,  the energy $\hbar \omega_k$ of the scalar field plays the role of the mass $m$ of the harmonic oscillator. The deformation parameter $\alpha$ can be then suitably redefined as
\be 
\tilde{\alpha}=
\beta\,\frac{\hbar^2 \omega_k^2}{2\,\mpl^2}=
2\,\beta\,\lp^2\,\omega_k^2
\ee
and the commutator between ladder operators becomes
\be
\label{modplanwav}
[A_k, A_{k'}^\dagger]
=
\frac{1}{1-\tilde{\alpha}}
\left[1
-
\tilde{\alpha}
\left(A_{k}^\dagger\, A_{k'}^\dagger
+A_{k}\, A_{k'}
-2\,A_{k}^\dagger\, A_{k'}
\right)
\right]
\delta(k-k')
\ .
\ee
\par
In Section~\ref{Quantization}, we have seen that the scalar field for an inertial observer can be quantized
both using plane-waves and boost-modes (see Eqs.~(\ref{eqn:planewavexpans})
and~(\ref{eqn:expansionfieldboost}), respectively).
In that context, the choice between these two representations is just a matter of convenience,
since the corresponding sets of ladder operators $a_k$ and $d_{\Omega}^{(\sigma)}$
are related by the canonical transformation~(\ref{eqn:operat-d}).
With deformed quantization rules, however, Lorentz invariance is violated and such an equivalence is not guaranteed.
Nevertheless, in the limit of very small deformation (that is, $\beta p^2\ll \mpl^2$), it appears reasonable to assume the same structure of the modified algebra for the two sets of operators. According to this argument, we thus conjecture the following deformation for the commutator in the boost-mode representation:
\begin{equation}
\label{eqn:d-D}
\left[D_{\Omega}^{(\sigma)}\,D_{\Omega'}^{(\sigma')\dagger}\right]
=
\frac{1}{1-\gamma}\left[
1-\gamma\,\left(
D_{\Omega}^{(\sigma)\dagger}\,D_{\Omega'}^{(-\sigma')\dagger}
+
D_{\Omega}^{(\sigma)}\,D_{\Omega'}^{(-\sigma')}
-D_{\Omega}^{(\sigma)\dagger}\,D_{\Omega'}^{(\sigma')}
-D_{\Omega}^{(-\sigma)\dagger}\,D_{\Omega'}^{(-\sigma')}
\right)
\right]
\delta_{\sigma\sigma'}\,
\delta(\Omega-\Omega')
\ ,
\end{equation}
where $D_{\Omega}^{\,(\sigma)}$ and $D_{\Omega}^{\,(\sigma)\dagger}$ are the ladder operators in the deformed algebra and the deforming parameter $\gamma$ is defined by
\be
\label{ips}
\gamma
=
\beta\,
\frac{\hbar^2 \omega^2}{2\,\mpl^2}
=
\beta\,
\frac{\hbar^2 a^2\,\Omega^2}{2\,\mpl^2}
= 2\,\beta\,\lp^2\,a^2\,\Omega^2
\ ,
\ee
being $\omega=a\Omega$ the Rindler frequency.
Some comments about Eq.~(\ref{eqn:d-D}) are needed.
First, in order to adapt the deformed commutator~(\ref{modplanwav}) to the boost operators $D$,
we have modified \emph{ad hoc} the definition of the deforming parameter $\tilde{\alpha}$
by replacing the plane-frequency $\omega_k$ with the  boost-mode frequency
$\omega=a\,\Omega$ [see Eq.~(\ref{ips})].
Furthermore, the commutator~(\ref{eqn:d-D}) has been multiplied by $\delta_{\sigma\sigma'}$
to ensure that the ladder operators in the right wedge $R_+$ are still commuting with the
corresponding operators in the left wedge $R_-$.
In addition, we symmetrized it with respect to $\sigma$ and $-\sigma$, so that
\begin{equation}
\label{eqn:equalcomm}
\left[D_{\Omega}^{(\sigma)},D_{\Omega'}^{(\sigma')\dagger}\right]
=
\left[D_{\Omega}^{(-\sigma)},D_{\Omega'}^{(-\sigma')\dagger}\right]
\ .
\end{equation}
By exploiting this property and recasting the Bogoliubov transformation~(\ref{eqn:bogotransform})
in the form
\begin{equation}
B_{\Omega}^{(\sigma)}
=
{\left[1+\mathcal{N}(\Omega)\right]}^{1/2}
\,D_{\Omega}^{(\sigma)}
+
{\mathcal{N}(\Omega)}^{1/2}\,
D_{\Omega}^{(-\sigma)\dagger}
\ ,
\label{eqn:newformbogotr}
\end{equation}
one can verify that the deformation~(\ref{eqn:d-D}) induces an identical modification to the
algebra of the Rindler operators $B$.
\par
GUP effects on the Unruh temperature can now be investigated by calculating the distribution
of $B$-quanta in the Minkowski vacuum $|0_{\rm M}\rangle$.
By use of the transformation~(\ref{eqn:newformbogotr}), it can be shown that
\begin{equation}
\label{eqn:modexpecnum}
\langle0_{\rm M}|\,
B_{\Omega}^{(\sigma)\dagger}\,B_{\Omega'}^{ (\sigma')}\,
|0_{\rm M}\rangle
=
\frac{1}{\left(e^{2\pi\Omega}-1\right)\left(1-\gamma\right)}\,
\delta_{\sigma\sigma'}\,
\delta(\Omega-\Omega')
\ ,
\end{equation}
to be compared with the standard Bose-Einstein distribution Eq.~(\ref{eqn:aspectval}).
As expected, the Unruh spectrum gets non-trivially modified by the deformed 
algebra~(\ref{eqn:d-D}) and loses its characteristic thermal behavior.
However, for Rindler frequencies $\Omega$ such that $\gamma\ll1$, namely (since $\beta\sim 1$) for $\hbar\omega\ll\mpl$, we have $e^{-\gamma}\simeq 1-\gamma$,
and Eq.~\eqref{eqn:modexpecnum} can be approximated as 
\begin{equation}
\label{eqn:approxexpecnum}
\langle0_{\rm M}|\,
B_{\Omega}^{(\sigma)\dagger}\,B_{\Omega'}^{ (\sigma')}\,
|0_{\rm M}\rangle
\simeq
\frac{1}{e^{2\pi\Omega-\gamma}-1}\,
\delta_{\sigma\sigma'}\,
\delta(\Omega-\Omega')
\ ,
\end{equation}
where we neglected the term linear in $\gamma$ in the denominator of the r.h.s.
We can interpret Eq.~(\ref{eqn:approxexpecnum}) as a shifted Bose-Einstein thermal distribution
by introducing a shifted Unruh temperature $T$ such that the term $(2\pi\Omega-\gamma)$ can be rewritten as
\be
\label{newT}
2\pi\Omega-\gamma 
\ = \
\frac{\hbar \,a\, \Omega}{k_{\rm B}\,T_{\rm U}} - \gamma
\ \equiv \
\frac{\hbar \,a\, \Omega}{k_{\rm B}\,T}
\ .
\ee
We thus find for the shifted Unruh temperature
\begin{equation}
T
=
\frac{T_{\rm U}}{1-\beta\,\pi\,\Omega\,k_{\rm B}^2\,T_{\rm U}^2/\mpl^2}
\simeq
T_{\rm U}\left(1 + \beta\, \pi\, \Omega \left(\frac{k_{\rm B}T_{\rm U}}{\mpl}\right)^2 \right)
=
T_{\rm U}
\left(1+\beta\,\pi\,\Omega\,\frac{\lp^2\,a^2}{\pi^2}\right)
\,.
\label{eqn:newT}
\end{equation}
We notice that such a modified temperature $T$ contains an explicit dependence on the Rindler frequency $\Omega$. This is due to the deformed structure of the
commutator (\ref{gupcomm}), which explicitly depends on $\hat{p}^2$, that is, essentially, on the energy of the considered quantum mode. So, it is not surprising to recover such an explicit dependence in the final formulae. Nevertheless, a simple thermodynamic argument allows us to get rid of this $\Omega$-dependance. In fact, for small deformations, we are still close to the thermal black body spectrum. Therefore the vast majority of the Unruh quanta will be emitted around a Rindler frequency $\omega$ such that $\hbar\,\omega\simeq k_{\rm B}\,T_{\rm U}$, which means $\Omega\approx 1/(2\pi)$.
For this typical frequency, Eq.~\eqref{eqn:newT} reproduces quite closely the heuristic 
estimate~\eqref{newTHeuristic}. In fact
\begin{equation}
T
\simeq
T_{\rm U}\left(1 + \frac{\beta}{2}  \left(\frac{k_{\rm B}T_{\rm U}}{\mpl}\right)^2 \right)
=
T_{\rm U}
\left(1 + \frac{\beta}{2}\frac{\lp^2\,a^2}{\pi^2}\right)
\ .
\end{equation}

It is also worth noting that the deformation of the algebra Eq.~(\ref{eqn:d-D}) would affect also the Hamiltonian. Therefore, the Rindler frequency $\Omega$ in Eq.~(\ref{newT}) should in principle be modified accordingly. In the present analysis, however, since we consider only small deformations of the quantization rules, we have reasonably neglected those corrections, thus approximating the modified Rindler Hamiltonian to the original one.

Concluding, for small deviations from the canonical quantization, we have found that the Unruh distribution maintains its original thermal spectrum, provided that a new temperature $T$ is defined as in Eq.~(\ref{eqn:newT}).
\section{Conclusions}
In the context of the Generalized Uncertainty Principle, we have computed the correction induced on the Unruh temperature by a deformed fundamental commutator. This has been done following two independent paths. First, we proceeded in a heuristic way, using very general and reasonable physical considerations. Already at this stage however we have been able to point out a dependence of the deformed Unruh temperature on the cubic power of the acceleration. These considerations have been substantiated and confirmed by means of a full fledged Quantum Field Theory calculation. This has been achieved by taking into account modified commutation relations for the ladder operators compatible with the GUP of Eq.(\ref{gupcomm}). In the limit of a small deformation of the commutator, we obtained again a dependence of the first correction term on the third power of acceleration. Besides, the more refined formalism of QFT has helped us to point out an explicit dependence of the deformed Unruh temperature from the Rindler frequency 
$\Omega$, which, on the other hand, was reasonably expected. A simple and effective thermodynamic argument has then been used to identify the values of most probable emission for the $\Omega$ Rindler frequency. As a consequence the QFT calculation matches in the end the heuristic estimate, indeed with almost the same numerical coefficients.
An avenue for further investigations could be the relation between the deviation from thermality of the Unruh radiation discussed in this paper and those found in different contexts (e.g. Ref.~\cite{Blasone:2017nbf}).


%
%
%
%
%
%

%
\end{document}